\documentclass[a4paper,11pt]{article}%
\usepackage{amsmath}
\usepackage{graphicx}%
\usepackage{amsfonts}%
\usepackage{amssymb}
\newtheorem{theorem}{Theorem}

\newtheorem{definition}[theorem]{Definition}

\newtheorem{proposition}[theorem]{Proposition}

\begin{document}

\title{A Unified Scheme for Generalized Sectors based on Selection Criteria. \\II. Spontaneously broken symmetries and some basic notions in quantum measurements}
\author{Izumi Ojima\\Research Institute for Mathematical Sciences, Kyoto University\\Kyoto 606-8502 Japan}
\date{}
\maketitle

\begin{abstract}
Continuing the analysis in a unified scheme for treating generalized
superselection sectors based upon the notion of selection criteria for states
of relevance in quantum physics, we extend the Doplicher-Roberts
superselection theory for recovering the field algebra and the gauge group (of
the first kind) from the data of group invariant observables to the situations
with spontaneous symmetry breakdown: in use of the machinery proposed, the
basic structural features of the theory with spontaneously broken symmetry,
are clarified in a clear-cut way, such as the degenerate vacua parametrized by
the variable belonging to the relevant homogeneous space, the Goldstone modes
and condensates.

\end{abstract}

\section{Introduction}

Continuuing the analysis started in the previous paper (called I) \cite{I}, we
apply here our unified method of treating generalized superselection sectors
based upon the notion of selection criteria to the situation with spontaneous
symmetry breakdown (SSB, for short). While the essence of the
\textit{superselection theory} of Doplicher-Haag-Roberts (DHR) \cite{DHR} and
of Doplicher-Roberts (DR) \cite{DR89, DR90} in algebraic quantum field theory
(QFT) \cite{Haag} was already explained briefly in I in a reformulated form
convenient for the present context, it may still be meaningful to mention some
of more general aspects of it for the sake of explaining the reason why we
regard our analysis of SSB as important.

The DHR-DR \textit{superselection theory} gives a general scheme for
understanding the relations between a symmetry and its observable consequences
in relativistic QFT. It tells us that, \textit{if }the internal symmetry of
the theory under consideration is, i) described by a gauge group $G$ of the
\textit{1st kind} (i.e., \textit{global} gauge symmetry), and is, ii)
\textit{unbroken}, the basic structure of the standard QFT can be recovered
totally from the data encoded in \textit{observables }$\mathfrak{A} $ which
are defined as $G$-invariant combinations of field operators (i.e.,
$\mathfrak{A}=\mathfrak{F}^{G}$: fixed-point subalgebra of the field algebra
$\mathfrak{F}$ under $G$) and constitute a \textit{net $\mathcal{O}$%
}$\longmapsto\mathfrak{A}(\mathcal{O})$ \textit{of local subalgebras of
observables} satisfying the\textit{\ local commutativity} (i.e., Einstein
causality). This implies, in particular, that the Bose/Fermi statistics of the
basic fields is automatically derived without necessity of introducing from
the outset \textit{unobservable} field operators such as \textit{fermionic
fields} subject to local \textit{anti}commutativity \textit{violating Einstein
causality}, which shows that they are simple mathematical devices for
bookkeeping of half-integer spin states. While all the non-trivial spacetime
behaviours are described here by the observable net \textit{$\mathcal{O}$%
}$\longmapsto\mathfrak{A}(\mathcal{O})$, the internal symmetry aspects are
encoded in the \textit{superselection structure}, which also originates from
the observable net.

The symmetry arising from this beautiful theory is, however, found always to
be \textit{unbroken},\textit{\ }excluding the situation of spontaneous
symmetry breakdown (SSB), which poses a question about the
``\textbf{stability}'' of this method, as remarked by the late Mosh\'{e} Flato
\cite{Flato96}. Indeed we know that many (actually, almost all) of the
``sacred symmetries'' in nature can be broken (explicitly or spontaneously) in
various situations: e.g.,

\begin{itemize}
\item SSB's of chiral symmetry in the electro-weak theory based upon
$SU(2)\times U(1)$, electromagnetic $U(1)$ in the superconductivity, and the
rotation symmetry $SO(3)$ in ferromagnetism, etc.,

\item Lorentz invariance is broken spontaneously in thermal equilibria with
$T\neq0^{\circ}K$ \cite{Oji86},

\item supersymmetry is shown to be unbroken only in the vacuum states
\cite{BO97}.
\end{itemize}

\noindent So, the question as to whether or not this theory can incorporate
systematically the cases of SSB is a real challenge to it, deserving serious
examination, and if the answer is yes, what kind of superselection structure
is realized in that case is a non-trivial interesting question. This sort of
investigation is expected also to give us some important clues for getting rid
of another restriction of \textit{global }gauge symmetries so as to
incorporate local gauge symmetries.

In the following, we give an affirmative answer to the above question,
revealing very interesting sector structures emerging from SSB: when we start
to extend this formalism to the situations with \textit{spontaneous symmetry
breakdown}, we encounter the presence of \textit{continuous sectors} (or,
``degenerate vacua'' in the physicist's terminology) parametrized by
continuous \textit{macroscopic order parameters}, as is seen in Sec.2.3. This
requires us to extend the traditional notion of sectors identified as the
discrete family of irreducible (or more generally, factorial) representations
of the observable algebra $\mathfrak{A}$ to incorporate the continuous ones.
What is more interesting physically and mathematically is the dual or
reciprocal relation found between the above degenerate vacua with classical
parameters and the quantum Goldstone modes found inside the sectors on a fixed
pure vacuum picked up from the degenerate vacua, because it leads us to the
point very close to such a physical expression that ``the Goldstone modes
search the degenerate vacua in a virtual way''. At the same time, this is also
related with the mathematical notion of \textit{duality} for homogeneous
spaces and their representations, as a natural extension of Tannaka-Krein
duality of compact groups. In this way, the basic structural features of the
theory with spontaneously broken symmetry, are clarified in a clear-cut way,
establishing mutual relationship among degenerate vacua, order parameters,
Goldstone modes and condenstates responsible for SSB.

As other examples of applications of the method, we examine in Sec.3 also some
basic notions supporting the physical and operational meanings of the
mathematical framework of quantum theory; the problem of physical
implementability of the probabilistic interpretations formulated and examined
in a form of the realizability and also that of state preparation as
reachability problem, in the general context of control theory, and so on.

\section{SSB-vacua as continuous sectors with order parameter whose quantum
precursor is Goldstone mode}

\subsection{Dual net $\mathfrak{A}^{d}$ and unbroken symmetry $H$}

To treat physically more interesting cases of \textbf{spontaneous symmetry
breakdown }(\textbf{SSB}),\textbf{\ }we need to extend the original DR
superselection theory where the internal symmetry is unbroken with unitary
implementers as long as the Haag duality $\mathfrak{A}^{d}(\mathcal{O}%
):=\pi_{0}(\mathfrak{A}(\mathcal{O}^{\prime}))^{\prime}=\pi_{0}(\mathfrak{A}%
(\mathcal{O}))$ (for $\mathcal{O}\in\mathcal{K}$) holds to play the crucial
roles. It can be shown that this property is also a necessary condition for
the field system with normal statistics and with unbroken symmetry (see,
\cite{DHR, DR90}). As pointed out by Roberts \cite{Roberts74}, SSB does not
take place without the breakdown of the Haag duality.

In the previous case with unbroken symmetry, the superselection sectors are
parametrized by the \textit{discrete} variables belonging to the dual
$\hat{G}$ of a compact group $G$. In the situation with SSB, one anticipates
physically the appearance of \textit{continuous\ }macroscopic \textit{order
parameters}, as typically exemplified by the continuous directions of
magnetization in the ferromagnetism, which strongly suggests the appearance of
\textit{continuous\ superselection sectors},\textit{\ }parametrized by
macroscopic order parameters\textit{. }This will be shown actually to be the
case in the following.

For the sake of convenience, we change the notation adopted in I; Sec.4 in the
unbroken symmetry case, so that the observable algebra $\mathfrak{A}$ and the
symmetry group $G$ in I;Sec.4 are replaced, respectively, by the dual net
$\mathfrak{A}^{d}$ (of the genuine observable algebra $\mathfrak{A}$) and the
group $H$ of \textit{unbroken remaining symmetry} in the present context. To
begin with, the correspondence between physically relevant states $\omega$
around the vacuum $\omega_{0}$ and such an endomorphism $\rho$ as
$\omega=\omega_{0}\circ\rho$ can be maintained when all the ingredients here
are understood in relation to the dual net $\mathfrak{A}^{d} $ under the
natural assumption of \textit{essential duality }%
\begin{equation}
\mathfrak{A}^{dd}=\mathfrak{A}^{d}%
\end{equation}
which is equivalent to the local commutativity of the dual net and is valid
whenever some Wightman fields are underlying the theory \cite{BDLR92}. First,
in view of the relation $\mathfrak{A}^{d}(\mathcal{O}^{\prime})^{\prime\prime
}=\mathfrak{A}(\mathcal{O}^{\prime})^{\prime\prime}$ \cite{Roberts74}, the
starting vacuum state and representation, $\omega_{0}$ and $(\pi
_{0},\mathfrak{H}_{0})$, can safely be extended from $\mathfrak{A}$ to
$\mathfrak{A}^{d}$ (meaning both the local net and the global algebra). Then
the DHR selection criterion is understood for the states $\omega$ of
$\mathfrak{A}^{d}$, as $\omega\upharpoonright_{\mathfrak{A}^{d}(\mathcal{O}%
^{\prime})}=\omega_{0}\upharpoonright_{\mathfrak{A}^{d}(\mathcal{O}^{\prime}%
)}$, and is equivalent to the existence of $\rho\in\mathcal{T}\subset
End(\mathfrak{A}^{d})$ such that $\omega=\omega_{0}\circ\rho$. On the basis of
these items, we can repeat the same procedure of constructing the field
algebra $\mathfrak{F}$ and the group $H$ of unbroken symmetry according to the
general method \cite{DR89, DR90}:
\begin{equation}
\mathfrak{F}=\mathfrak{A}^{d}\underset{\mathcal{O}_{d_{0}}^{H}}{\otimes
}\mathcal{O}_{d_{0}},\text{ \ \ \ }H=Gal(\mathfrak{F/A}^{d}).
\end{equation}

\subsection{Spontaneously broken symmetry}

What we are going to show is the following superselection structure of the
theory with spontaneously broken symmetry described by the Galois group
$G:=Gal(\mathfrak{F}/\mathfrak{A})\supset H$. First we consider the
irreducible $H$-covariant vacuum representation $(\pi,U,\mathfrak{H})$ of the
system $\mathfrak{F}\underset{\tau}{\curvearrowleft}H$, $\pi(\tau
_{h}(F))=U(h)\pi(F)U(h)^{\ast}$ for $\forall F\in\mathfrak{F}$, $\forall h\in
H$, containing the original representation $(\pi_{0},\mathfrak{H}_{0})$ of
$\mathfrak{A}$ and $\mathfrak{A}^{d}$ as the cyclic fixed-point subspace under
$U(H)$: $\mathfrak{H}_{0}=\{\xi\in\mathfrak{H};$ $U(h)\xi=\xi$ for $\forall
h\in H\}$, $\overline{\pi(\mathfrak{F)H}_{0}}=\mathfrak{H}$. Then, \ according
to the DHR sector structure in the unbroken case \cite{DHR}, we have%
\begin{equation}
\mathfrak{Z}(\pi(\mathfrak{A}^{d})^{\prime\prime})=\underset{\eta\in
\hat{H}}{\oplus}\mathbb{C}(\mathbf{1}_{\mathfrak{H}_{\eta}}\otimes
\mathbf{1}_{W_{\eta}})=C(\hat{H})=\mathfrak{Z}(U(H)^{\prime\prime}).
\end{equation}
Since this group $H$ is the maximal group of unbroken symmetry in the
irreducible vacuum situation, the group $G$ bigger than $H$ cannot be
unitarily implemented in the above representation $(\pi,\mathfrak{H})$ of
$\mathfrak{F}$, which is just the precise meaning of the SSB of $G$ in the
present situation. In more general situations the precise definition can be
given by:

\begin{definition}
\label{Def:SSB}A symmetry described by a (strongly continous) automorphic
action $\tau$ of $G$ on the field algebra $\mathfrak{F}$ is said to be
\textbf{unbroken }in a given representation $(\pi,\mathfrak{H})$ of
$\mathfrak{F}$ if each \textbf{factor} subrepresentation $(\sigma
,\mathfrak{H}_{\sigma})$, $\sigma(\mathfrak{F)}^{\prime}\cap\sigma
(\mathfrak{F)}^{\prime\prime}=\mathbb{C}\mathbf{1}_{\mathfrak{H}_{\sigma}}$,
appearing in the central decomposition of $(\pi,\mathfrak{H})$ admits a
\textbf{covariant representation} of the system $G\overset{\tau}%
{\curvearrowright}\mathfrak{F}$ in the sense that there exists a (strongly
continuous) unitary representation $(U_{\sigma},\mathfrak{H}_{\sigma})$ of $G$
verifying the relation $\sigma(\tau_{g}(F))=U_{\sigma}(g)\sigma(F)U_{\sigma
}(g)^{\ast}$ for $\forall g\in G,\forall F\in\mathfrak{F}$. If the symmetry is
not unbroken, it is said to be \textbf{broken spontaneously}.
\end{definition}

Note that the above characterization of unbroken symmetry can be reformulated
equivalently as the pointwise invariance of the spectrum of centre of
$\pi(\mathfrak{F)}^{\prime\prime}$ under $G$. Therefore, SSB\ means in short
the \textit{conflict between the unitary implementability and the factoriality
(=triviality of centres)} \cite{IO99}. The situation with SSB is seen to
exhibit the features of the so-called \textit{``infrared
instability''\textbf{\ }}under the action of $G$, because $G$ does not
stabilize the spectrum of centre which can be viewed physically as
\textit{macroscopic order parameters }emerging in the infrared (=low energy)
regions. Since the above definition of SSB still allows the mixture of
unbroken and broken \textit{sub}representations of a given $\pi$, we need to
decompose the spectrum of the centre of $\pi(\mathfrak{F)}^{\prime\prime}$
into domains each of which is \textit{ergodic} under $G$ (central ergodicity).
Then, $\pi$ is decomposed into the direct sum (or, direct integral) of
unbroken factor representations and broken non-factor representations, each
component of which is stable under $G$. Thus we obtain a \textit{phase diagram
}on the spectrum of the centre.

As indicated above, the intuitive physical picture of \textit{order parameters
arising from the SSB} from $G$ down to $H$ is realized in connection with the
sector structure of the whole theory involving the presence of
\textit{continuous sectors} parametrized by $\dot{g}\in H\backslash G$. Here
we need to combine the above two formulations of discrete sectors of unbroken
internal symmetry (I;Sec.4) and of continuous sectors (I;Sec.2) in the
following way. One important point to be mentioned here is that our motivation
for treating here the \textit{centre}s at various levels of representations is
always coming from the natural and inevitable occurrence of \textit{disjoint}
representations which leads to the appearance of \textit{macroscopic order
parameters to classify different modes of }macroscopic manifestations of
microscopic systems; this should be properly contrasted to a mathematical
pursuit of generalizing the pre-existing machinery involving factor algebras
to non-factorial ones.

According to this formulation, we should find such a covariant representation
of the system $(\mathfrak{F}\underset{\tau}{\curvearrowleft}G)$ as
implementing \textit{minimally} the broken $G$ in the sense of \textit{central
ergodicity }under $G$(, which is implied by the factoriality of representation
of the crossed product algebra $\mathfrak{F}\underset{\tau}{\rtimes}G$ of
$\mathfrak{F}$ with $G$). Since the subgroup $H$ is unbroken in the
irreducible covariant representation $(\pi,U,\mathfrak{H)}$ of $\mathfrak{F}%
\underset{\tau}{\curvearrowleft}H$, what we seek for can actually be provided
by the representation $(\hat{\pi},\mathfrak{\hat{H}})$, induced from
$(\pi,U,\mathfrak{H)}$, of the crossed product $\mathfrak{\hat{F}%
}:=\mathfrak{F}\underset{\hat{\tau}}{\rtimes}(H\backslash G)=\Gamma
(G\times_{H}\mathfrak{F})$ of $\mathfrak{F}$ with the homogeneous space
$H\backslash G$ (having the right $G$-action being transitive, and hence,
trivially $G$-ergodic), which can be identified with the algebra of
$H$-equivariant norm-continuous functions $\hat{F}:G\rightarrow\mathfrak{F}$,
\begin{equation}
\hat{F}(hg)=\tau_{h}(\hat{F}(g)).\label{equivA}%
\end{equation}
Denoting $d\xi$ the left-invariant Haar measure on $G/H$ (equipped with the
\textit{left }$G$-action), we define a Hilbert space $\mathfrak{\hat{H}}$ of
$L^{2}$-sections of $\Gamma(G\times_{H}\mathfrak{H})$ by%
\begin{equation}
\mathfrak{\hat{H}}=\int_{\xi\in G/H}^{\oplus}(d\xi)^{1/2}\mathfrak{H}%
=L^{2}(\Gamma(G\times_{H}\mathfrak{H}),d\xi),
\end{equation}
which can also be identified with the $L^{2}$-space of $\mathfrak{H}$-valued
$H$-equivariant functions $\psi$ on $G$,
\begin{equation}
\psi(gh)=U(h^{-1})\psi(g).\label{equivS}%
\end{equation}
On this $\mathfrak{\hat{H}}$, a representation of $\mathfrak{\hat{F}}$ is
defined by
\begin{equation}
(\hat{\pi}(\hat{F})\psi)(g)=\pi(\hat{F}(g^{-1}))(\psi(g))\text{ \ \ \ for
}\hat{F}\in\mathfrak{\hat{F}}\text{, }\psi\in\mathfrak{\hat{H}}\text{, }g\in
G,\label{crss-rep}%
\end{equation}
which is compatible with the above equivariance condition, (\ref{equivS}):
\begin{align}
& (\hat{\pi}(\hat{F})\psi)(gh)=\pi(\hat{F}(h^{-1}g^{-1}))(\psi(gh))\nonumber\\
& =U(h^{-1})\pi(\hat{F}(g^{-1}))U(h)U(h^{-1})(\psi(g))=U(h^{-1})(\hat{\pi
}(\hat{F})\psi)(gh).
\end{align}
As is well-known, this representation $(\hat{\pi},\mathfrak{\hat{H}})$ is
equivalent to the covariant representation $(\bar{\pi}%
,\hat{U},\mathfrak{\hat{H}})$ of the dynamical system $\mathfrak{F}%
\underset{\tau}{\curvearrowleft}G$ defined on $\mathfrak{\hat{H}}$ by
$(\hat{U}(g)\psi)(g_{1}):=\psi(g^{-1}g_{1})$, $(\bar{\pi}(F)\psi)(g):=\pi
(\tau_{g^{-1}}(F))\psi(g)$ ($\psi\in\mathfrak{\hat{H}}$) and satisfying
$\bar{\pi}(\tau_{g}(F))=\hat{U}(g)\bar{\pi}(F)\hat{U}(g)^{-1}$, through the
relations $\hat{\pi}:=\bar{\pi}\rtimes\hat{U}$ (meaning $\hat{\pi}(\hat
{F})=\int dg\ \bar{\pi}(\hat{F}(g))\hat{U}(g)$) and $\hat{U}(f):=\hat{\pi
}(\hat{f})$ with $\hat{f}:G\ni g\longmapsto f(g)\mathbf{1}\in\mathfrak{F}$ for
$f\in C(G)$ and $\bar{\pi}(F):=s-\lim_{\alpha\rightarrow\infty}\hat{\pi
}(F_{\alpha})$, $F_{\alpha}(g)=Ff_{\alpha}(g)$, $f_{\alpha}\rightarrow
\delta_{e}$, $f_{\alpha}\in C(H\backslash G)$. Here also, all the operations
are compatible with the constraints of $H$-equivariance.

\subsection{Sector structures and \textit{c}$\rightarrow$\textit{q} channel}

What is important for us here is to clarify the \textit{centre} of
$\mathfrak{A}$ in the representation $(\hat{\pi}\upharpoonright_{\mathfrak{A}%
},\mathfrak{\hat{H}})$ obtained by the restriction of $(\hat{\pi
},\mathfrak{\hat{H}})$ from $\mathfrak{\hat{F}}$ to $\mathfrak{A}$, which givs
the superselection sectors as seen below. The mutual relations among the
relevant C*-algebras can be summarized in the following commuting diagram:
\[%
\begin{array}
[c]{ccccc}
&  & \mathfrak{\hat{F}}=\Gamma(G\underset{H}{\times}\mathfrak{F}) &  & \\
& ^{^{\hat{m}_{G}}}\swarrow\nearrow_{\hat{\imath}_{G}} & \circlearrowright &
_{\hat{\imath}_{H\backslash G}}\nwarrow\searrow^{\hat{m}_{H\backslash G}} & \\
\mathfrak{\hat{F}}^{G}=\mathfrak{A}^{d}=\mathfrak{F}^{H} & \leftrightarrows &
\overset{m_{H}}{\underset{i_{H}}{\leftrightarrows}} & \leftrightarrows &
\mathfrak{F}=\mathfrak{A}^{d}\underset{\mathcal{O}_{d_{0}}^{H}}{\otimes
}\mathcal{O}_{d_{0}}\\
& _{m_{G/H}}\searrow\nwarrow^{i_{G/H}} & \circlearrowright & ^{i_{G}}%
\nearrow\swarrow_{m_{G}} & \\
&  & \mathfrak{A}=\mathfrak{F}^{G} &  &
\end{array}
,
\]
where the maps $i_{G}$ and $m_{G}$, etc. are, respectively, the embedding maps
(of a C*-algebra into another) and the conditional expectations, such as
\begin{equation}
m_{G}:\mathfrak{F}\ni F\longmapsto m_{G}(F):=\int_{G}dg\tau_{g}(F)\in
\mathfrak{A.}%
\end{equation}
It is interesting to note that the relation between $\mathfrak{\hat{F}}$ and
$\mathfrak{A}^{d}$ is just parallel to that between $\mathfrak{F}$ and
$\mathfrak{A}$, which becomes relevant later to our discussion of the sector
structures. Using the relations, $\hat{\pi}\upharpoonright_{\mathfrak{A}%
}(\mathfrak{A})=\int_{\dot{g}\in H\backslash G}^{\oplus}d\dot{g}$
$\pi\upharpoonright_{\mathfrak{A}}(\mathfrak{A})=C(H\backslash G)\otimes
\pi\upharpoonright_{\mathfrak{A}}(\mathfrak{A})$ and $\pi\upharpoonright
_{\mathfrak{A}^{d}}(\mathfrak{A}^{d})=\oplus_{\eta\in\hat{H}}(\pi_{\eta
}(\mathfrak{A}^{d})\otimes\mathbf{1}_{W_{\eta}})$, we can show
\begin{align}
\mathfrak{Z}(\hat{\pi}\!\upharpoonright\!_{\mathfrak{A}}(\mathfrak{A}%
)^{\prime\prime})  & =\left\{
\begin{array}
[c]{c}%
\int_{\dot{g}\in H\backslash G}^{\oplus}d\dot{g}\ [w-\lim_{\alpha
\rightarrow\infty}\pi(A_{\alpha})];\text{ net }\{A_{\alpha}\}\subset
\mathfrak{A},\\
\text{ \ \ \ \ \ \ \ \ \ \ \ \ \ s.t.\ }\mathfrak{[}w-\lim_{\alpha
\rightarrow\infty}\pi\upharpoonright_{\mathfrak{A}}(A_{\alpha})]\in
\mathfrak{Z}(\pi(\mathfrak{A})^{\prime\prime})
\end{array}
\right\} \nonumber\\
& =\left\{
\begin{array}
[c]{c}%
\int_{\dot{g}\in H\backslash G}^{\oplus}d\dot{g}\ [\oplus_{\eta\in
\hat{H}}[w-\lim_{\alpha\rightarrow\infty}(\pi_{\eta}\upharpoonright
_{\mathfrak{A}}(A_{\alpha})\otimes\mathbf{1}_{W_{\eta}})]];\\
\text{net }\{A_{\alpha}\}\subset\mathfrak{A}\text{\ s.t.\ \ }\mathfrak{[}%
w-\lim_{\alpha\rightarrow\infty}\pi_{\eta}\upharpoonright_{\mathfrak{A}%
}(A_{\alpha})]\\
\in\mathfrak{Z}(\pi_{\eta}\upharpoonright_{\mathfrak{A}}(\mathfrak{A}%
)^{\prime\prime})
\end{array}
\right\} \nonumber\\
& =L^{\infty}(H\backslash G;d\dot{g})\otimes\lbrack\oplus_{\eta\in
\hat{H}}\mathfrak{Z}(\pi_{\eta}\upharpoonright_{\mathfrak{A}}(\mathfrak{A}%
)^{\prime\prime})].
\end{align}

On the basis of these structures of relevant centres of representations, we
define a \textit{c}$\rightarrow$\textit{q} \textit{channel} $\Psi$ as
follows:
\begin{align}
& \Psi:\mathfrak{A}^{d}\ni B\longmapsto\Psi(B)\in C(H\backslash G)\otimes
\lbrack\oplus_{\chi\in\hat{H}}\mathfrak{Z}(\pi_{\chi}(\mathfrak{A}%
)^{\prime\prime})],\nonumber\\
& \Psi(B):(H\backslash G)\times Spec(\oplus_{\chi\in\hat{H}}\mathfrak{Z}%
(\pi_{\chi}(\mathfrak{A})^{\prime\prime}))\ni(\dot{g},(\eta,\gamma
))\longmapsto\nonumber\\
& \qquad\qquad\rightarrow(\omega_{0}\circ\sigma_{\gamma}\circ m_{G/H}\circ
\rho_{\eta}\circ m_{H})(\tau_{\dot{g}}(B)).
\end{align}
Here, $\rho_{\eta}\in\mathcal{T}$ is a local endomorphism of $\mathfrak{A}%
^{d}$ belonging to the DR-category $\mathcal{T}_{\mathfrak{A}^{d}}%
\mathcal{\ }$on $\mathfrak{A}^{d}$ and $\gamma\in Spec(\mathfrak{Z}(\pi_{\eta
}\upharpoonright_{\mathfrak{A}}(\mathfrak{A})^{\prime\prime}))$ is an
irreducible representation $\gamma\in\hat{G}$ of $G$ whose restriction to $H$
contains $\eta\in\hat{H}$: i.e., $\exists\eta_{1}\in RepH$ s.t. $\gamma
\upharpoonright_{H}=\eta\oplus\eta_{1}$. (Although not fully expressed in
either of the above formulae, nor becomes relevant to the context of
discussing \textit{c}$\rightarrow$\textit{q} \textit{channel} $\Psi$, we need
to be careful about such constraints imposed on the components of the
\textit{central elements} that all the components $c_{\gamma}$'s belonging to
the same $\gamma\in\hat{G}$ should be identical even if they appear as
subrepresentations $\pi_{\gamma}(\mathfrak{A})$ of $\pi_{\eta}\upharpoonright
_{\mathfrak{A}}$'s which are \textit{mutually disjoint as representations of}
$\mathfrak{A}^{d}$: i.e., $c\in\oplus_{\chi\in\hat{H}}\mathfrak{Z}(\pi_{\chi
}(\mathfrak{A})^{\prime\prime})$ can be decomposed as $c=(c_{\eta})_{\eta
\in\hat{H}}$, $c_{\eta}=(c_{\gamma}^{(\eta)})_{\gamma\in Spec(\mathfrak{Z}%
(\pi_{\eta}\upharpoonright_{\mathfrak{A}}(\mathfrak{A})^{\prime\prime})}$,
where $c_{\eta_{1}}$ and $c_{\eta_{2}}$ for $\eta_{1},\eta_{2}\in\hat{H} $,
$\eta_{1}\neq\eta_{2}$, are independent up to the constraint that $c_{\gamma
}^{(\eta_{1})}=c_{\gamma}^{(\eta_{2})}$ for any $\gamma\in\hat{G}$).

Before a field algebra $\mathfrak{F}$ is constructed, the Doplicher-Roberts
method based on the local endomorphisms and the related Cuntz algebras
\cite{Cuntz} seems to be the\ only possible path from $\mathfrak{A}$ to the
pair of $\mathfrak{F}$ and $G$ \textit{without knowing} either of them, which
has necessarily led us to an \textit{unbroken }and \textit{compact }symmetry
group. However, in the situation of SSB with massless spectrum, there is
\textit{no} reason \textit{nor }guarantee for the broken group
$G=Gal(\mathfrak{F}/\mathfrak{A}) $ to be compact, as shown in \cite{BDLR92}
through the counter-examples. Fortunately, once $\mathfrak{F}$ is so
constructed from the dual net $\mathfrak{A}^{d}$ and the DR category
$\mathcal{T}_{\mathfrak{A}^{d}}$ as to show certain kinds of stability
properties (as will be discussed later), we need not any more stick to the
original line of thought inherent to the Doplicher-Roberts theory: having at
hand the information on the group $G=Gal(\mathfrak{F}/\mathfrak{A})$, we can
control the mutual relations among $\mathfrak{F}$, $G$, $\mathfrak{A}^{d}$ and
$\mathfrak{A}$ by means of various versions of crossed products applicable to
$G$, \textit{irrespectively} of whether it is compact or not \cite{Nak-Take}.
However, when $G=Gal(\mathfrak{F}/\mathfrak{A})$ is compact, as is common in
the physical examples of SSB (such as the case of chiral $SU(2)\times SU(2)$
down to the vectorial $SU(2)$), we can get more detailed information as
follows along the line suggested in \cite{DR90}, in full use of the
corresponding category $\mathcal{T}_{\mathfrak{A}}$ on $\mathfrak{A}$. First,
we have the relation
\begin{align}
& Spec(\mathfrak{Z}(\pi_{\eta}\upharpoonright_{\mathfrak{A}}(\mathfrak{A}%
)^{\prime\prime}))\nonumber\\
& =\{\gamma\in\hat{G};\pi_{\gamma}=\pi_{0}\circ\sigma_{\gamma}\text{:
subrepresentation of }\pi\circ\lbrack\rho_{\eta}]\upharpoonright
_{\mathfrak{A}}\},
\end{align}
where $[\rho_{\eta}]$ is a minimal $\rho\in\mathcal{T}_{\mathfrak{A}^{d}} $
s.t. $\rho(\mathfrak{A})\subset\mathfrak{A}$ and $\rho\succ\rho_{\eta}$ w.r.t.
the ordering defined by
\begin{align}
\rho\succ\rho_{1}  & \overset{\text{def.}}{\Longleftrightarrow}\exists\rho
_{2}\in\mathcal{T}_{\mathfrak{A}^{d}}\exists w_{1},w_{2}:\text{isometries}%
\in\mathfrak{A}^{d}\nonumber\\
& \text{s.t. }\rho(B)=w_{1}\rho_{1}(B)w_{1}^{\ast}+w_{2}\rho_{2}(B)w_{2}%
^{\ast}\text{ for }B\in\mathfrak{A}^{d},
\end{align}
which is equivalent to $\pi_{\rho}=\pi_{\rho_{1}}\oplus\pi_{\rho_{2}}$ as
representations of $\mathfrak{A}^{d}$ and to $\eta_{\rho}=\eta_{\rho_{1}%
}\oplus\eta_{\rho_{2}}$ as representations of $H$. $\sigma_{\gamma}$ is a
local endomorphism of $\mathfrak{A}$ corresponding to $\gamma\in\hat{G}$,
which means that $Spec(\mathfrak{Z}(\pi_{\eta}\upharpoonright_{\mathfrak{A}%
}(\mathfrak{A})^{\prime\prime}))$ consists of the $G$-charges $\gamma
\in\hat{G}$ \textit{contained in an }$H$\textit{-sector} having $\eta
\in\hat{H}$ among its components. In contrast to the analysis \cite{Roberts74,
ILP} in the opposite direction from the data of a given family of subspaces in
the representations of $G$ to determine the corresponding subgroup $H$ in $G$,
here we need to examine the sector structures from the smaller group $H$ to a
bigger one $G$. To be more precise, we can prove and utilize the following
results under the assumption of the compactness of $G=Gal(\mathfrak{F}%
/\mathfrak{A})$:

\begin{enumerate}
\item \textit{Finite-dimensional induction} for a compact pair $H$
$\hookrightarrow G$: \newline Any finite-dimensional unitary representation
$(\eta,W)$ of $H$ can be extended to a representation $(\gamma,V)$ of $G$ by
taking a direct sum $\gamma|_{H}\cong\eta\oplus\eta^{\prime}$ with a suitable
representation $(\eta^{\prime},W^{\prime})$ of $H$ (for proof, see
\cite{TodaMimura}). At the level of a field algebra, this kind of induction is
sufficient, in contrast to the situations of states for which the genuine
Mackey induction is indispensable.

\item Stability and consistency of field algebra construction in SSB:
\newline In use of the above result, one can verify the \textit{stability} of
the crossed product construction of the field algebra under the change of
Cuntz algebras as the isomorphism between $\mathfrak{F}$ due to the original
DR construction from the dual net $\mathfrak{A}^{d}$ and the crossed product
of $\mathfrak{A}^{d}$ with a Cuntz algebra $\mathcal{O}_{d}$ for any $d>d_{0}%
$:
\begin{equation}
\mathfrak{F}:=\mathfrak{A}^{d}\underset{\mathcal{O}_{d_{0}}^{H}}{\otimes
}\mathcal{O}_{d_{0}}\cong\mathfrak{A}^{d}\underset{\mathcal{O}_{d}^{H}%
}{\otimes}\mathcal{O}_{d},
\end{equation}
where the isomorphism $\cong$ is due to a joint work \cite{NO} with T. Nozawa.
While the relation $g(\mathfrak{A}^{d})=\mathfrak{A}^{d}=\mathfrak{F}^{H}$ for
$g\in G$ requires $g\in N_{H}$, the normalizer of unbroken $H$ in
$G=Gal(\mathfrak{F}/\mathfrak{A})$, the equality
\begin{equation}
g(\mathfrak{A}^{d})\underset{\mathcal{O}_{d}^{gHg^{-1}}}{\otimes}%
\mathcal{O}_{d}=g(\mathfrak{A}^{d}\underset{\mathcal{O}_{d}^{H}}{\otimes
}\mathcal{O}_{d})=\mathfrak{F}%
\end{equation}
can be verified even for such $g\in G$ that $g\notin$ $N_{H}$, which shows the
\textit{consistency} of the construction method with the action of $G$ bigger
than $H$. \newline \ \ \ \ \ While the relation $Gal(\mathfrak{A}%
^{d}/\mathfrak{A})=N_{H}/H$ was verified in \cite{BDLR93}, their analysis of
degenerate vacua was restricted only to $N_{H}$ in order to avoid
$g(\mathfrak{A}^{d})\neq\mathfrak{A}^{d}$. In the physically interesting
situations involving Lie groups, however, the reductivity of a compact Lie
group $H$ $\hookrightarrow G$ implies that $N_{H}/H$ is abelian and/or
discrete with a vanishing Lie brackets, which does not seem to be relevant to
the physically meaningful contexts.\newline 

\item \textit{Duality for homogeneous spaces} and its endomorphism version:
For a compact group pair $H\hookrightarrow G$, the definition of $Rep_{G/H}$
and the mutual relations among $Rep_{G}$, $Rep_{H}$ and $Rep_{G/H}$ can be
described in terms of a \textit{homotopy-fibre category} $Rep_{G}$ over
$Rep_{H}$ with $Rep_{G/H}$ as homotopy fibre (S. Maumary \cite{Maum}): Over
$\eta\in Rep_{H}$ a homotopy fibre (h-fibre for short) is given by a category
$\eta/Rep_{G}$ (which is called a comma category under $\eta$ \cite{MacL}
whose objects are pairs $(\gamma,T)$ of $\gamma\in$ $Rep_{G}$ and $T\in
Rep_{H}(\eta,\gamma|_{H})$ and whose morphisms $\phi:(\gamma,T)\rightarrow
(\gamma^{\prime},T^{\prime})$ are given by $\phi\in Rep_{G}(\gamma
,\gamma^{\prime})$ s.t. $T^{\prime}=\phi\circ T$:
\begin{equation}%
\begin{array}
[c]{ccccc}
&  & \eta &  & \\
& \swarrow_{T} &  & _{T^{\prime}}\searrow & \\
\gamma|_{H} & \rightarrow & \underset{\phi|_{H}}{\rightarrow} & \rightarrow &
\gamma^{\prime}|_{H}\\
i_{H}\uparrow\text{\ \ } &  &  &  & \text{ \ }\uparrow i_{H}\\
\gamma & \rightarrow & \underset{\phi}{\rightarrow} & \rightarrow &
\gamma^{\prime}%
\end{array}
\end{equation}
(To be more precise, the comma category $\eta/Rep_{G}$ is to be understood as
$\eta/i_{H}$ where the functor $i_{H}:Rep_{G}\rightarrow Rep_{H}$ is the
restriction of $G$-representations to the subgroup $H$ of $G$.)\newline
\ \ \ \ The h-fibre over the trivial representation $\eta=\iota\in Rep_{H}$ of
$H$ is nothing but the category of \textit{linear representations of }$G/H $
due to Iwahori-Sugiura \cite{IwaSug}, to which any other h-fibres can be shown
to be homotopically equivalent \cite{Maum}.\newline \ \ \ \ The version in
terms of endomorphisms dual to the above h-fibre category is given as follows
\cite{Oji2002}: \newline 

$End(\mathfrak{A}^{d})$ $\ \ \ \supset\mathcal{T}_{\mathfrak{A}^{d}%
}\ \ \ \ \ \ \ \ \ \ \ \ \ \ \ \ \ \ \ \ \ \ \ \ \ \ \ \ \ \longleftrightarrow
Rep_{H} $

$End(\mathfrak{A}^{d},\mathfrak{A})\supset\mathcal{T}_{\mathfrak{A}}=\{\rho
\in\mathcal{T}_{\mathfrak{A}^{d}};\rho(\mathfrak{A)\subset A}%
\}\longleftrightarrow Rep_{G}$\medskip

\ \ \ $\ \ $ \ \ \ \ \ \ \ \ \ \ \ \ $\mathcal{T}_{\mathfrak{A}^{d}%
}/\mathcal{T}_{\mathfrak{A}}$%
\ \ \ \ \ \ \ \ \ \ \ \ \ \ \ \ \ \ \ \ \ \ \ \ \ \ $\longleftrightarrow
Rep_{G/H}$\newline The h-fibre category over $\rho\in\mathcal{T}%
_{\mathfrak{A}^{d}}$ is given by $\rho/\mathcal{T}_{\mathfrak{A}}$ [or, more
precisely, $\rho/\mathcal{D} $ with $\mathcal{D}$ being the functor
$\mathcal{T}_{\mathfrak{A}}\ni\sigma\longmapsto\tilde{\sigma}\in
\mathcal{T}_{\mathfrak{A}^{d}}$ extending endomorphisms from $\mathfrak{A}$ to
$\mathfrak{A}^{d}$] with the object set
\begin{equation}
\{(\sigma,T);\sigma\in\mathcal{T}_{\mathfrak{A}},T\in(\rho,\tilde{\sigma
})\subset\mathfrak{A}^{d}\}
\end{equation}
and with the set of morphisms
\begin{equation}
\{\phi:(\sigma,T)\rightarrow(\sigma^{\prime},T^{\prime});\phi\in
\mathcal{T}_{\mathfrak{A}}(\sigma,\sigma^{\prime})\subset\mathfrak{A}%
,T^{\prime}=\phi\circ T\in(\rho,\tilde{\sigma}^{\prime})\subset\mathfrak{A}%
^{d}\}
\end{equation}
(: semidirect product of $\mathcal{T}_{\mathfrak{A}}$ and $\mathfrak{A}^{d}$),
where
\begin{equation}
\tilde{\sigma}=\sigma\circ j_{\mathfrak{A}(\mathcal{O})}^{\prime}\circ
j_{\mathfrak{A}^{d}(\mathcal{O})}%
\end{equation}
gives the extension of endomorphism $\sigma\in\mathcal{T}_{\mathfrak{A}}$ of
$\mathfrak{A}(\mathcal{O})$ to $\mathfrak{A}^{d}(\mathcal{O})$ for
$\mathcal{O}=$supp$\sigma$ and $j_{\mathfrak{A}(\mathcal{O})}^{\prime}$,
$j_{\mathfrak{A}^{d}(\mathcal{O})}$ are the modular conjugations of von
Neumann algebras $\mathfrak{A}(\mathcal{O})^{\prime}$ and $\mathfrak{A}%
^{d}(\mathcal{O})$, respectively:
\begin{equation}
\mathfrak{A}^{d}(\mathcal{O})\overset{j_{\mathfrak{A}^{d}(\mathcal{O})}%
}{\rightarrow}\mathfrak{A}^{d}(\mathcal{O})^{\prime}\subset\mathfrak{A}%
(\mathcal{O})^{\prime}\overset{j_{\mathfrak{A}(\mathcal{O})}^{\prime}%
}{\rightarrow}\mathfrak{A}(\mathcal{O})^{\prime\prime}=\mathfrak{A}%
(\mathcal{O}).
\end{equation}
\newline \ \ \ \ \ Corresponding to $\mathcal{T}_{\mathfrak{A}}\overset
{i}{\hookrightarrow}\mathcal{T}_{\mathfrak{A}^{d}}\overset{V}{\hookrightarrow
}Hilb$, we have an embedding map $H=End_{\otimes}(V)\overset{j}%
{\hookrightarrow}End_{\otimes}(V\circ i)\equiv G$, as a result of which the
bigger group $G$ suffering from SSB is determined. \newline $\because)$ For
any $u\in H=End_{\otimes}(V)$, $\forall\rho\in\mathcal{T}_{\mathfrak{A}^{d}},$
$\exists u_{\rho}:V_{\rho}\rightarrow V_{\rho}$ s.t. for $\forall T\in
(\rho_{1},\rho_{2})$ $V_{T}\circ u_{\rho_{1}}=u_{\rho_{2}}\circ V_{T}$. Then,
for any $\sigma\in\mathcal{T}_{\mathfrak{A}}$, $i(\sigma)=\tilde{\sigma}%
\in\mathcal{T}_{\mathfrak{A}^{d}}$ and $\forall S\in(\sigma_{1},\sigma_{2})$
$i(S)\in(i(\sigma_{1}),i(\sigma_{2}))\subset\mathfrak{A}\subset\mathfrak{A}%
^{d},$ $V_{i(S)}\circ u_{i(\sigma_{1})}=u_{i(\sigma_{2})}\circ V_{i(S)}$,
which means $j(u)=u_{i(\cdot)}:\mathcal{T}_{\mathfrak{A}}\rightarrow
\mathcal{U}(V_{i(\cdot)})$ is a natural unitary transformation from the
functor $V\circ i=i^{\ast}(V)$ to itself, belonging to $End_{\otimes}(V\circ
i)=G$. \newline \ \ \ \ \ Then, for each $\sigma\in\mathcal{T}_{\mathfrak{A}}$
, we obtain $\gamma_{\sigma}|_{H}=\gamma_{\sigma}\circ j=\eta_{i(\sigma)}$,
which states that for each $H$-representation of the form $\eta_{i(\sigma)}$
($\sigma\in\mathcal{T}_{\mathfrak{A}}$), there is a $G$-representation
$\gamma_{\sigma}$ whose restriction to $H$ is $\eta_{i(\sigma)}=\gamma
_{\sigma}|_{H}$. This is just the categorical dual formulation of the
finite-dimensional induction in 1. \newline 

\item Generalizing a theorem in \cite{DR89}, we obtain \cite{Oji2002}

\begin{proposition}
If $\mathfrak{Z}(\mathfrak{A}^{d})=\mathbb{C}\mathbf{1}$ (as a C*-algebra), we
have the following relations \newline
\begin{align}
& \mathfrak{A}^{d}\underset{\mathcal{O}_{d}^{G}}{\otimes}\mathcal{O}_{d}
=\Gamma(G\underset{H}{\times}(\mathfrak{A}^{d}\underset{\mathcal{O}_{d}^{H}%
}{\otimes}\mathcal{O}_{d}))=:\mathfrak{\hat{F}},\\
& \mathfrak{\hat{F}}^{G} =\mathfrak{A}^{d}=(\mathfrak{A}^{d}\underset
{\mathcal{O}_{d}^{H}}{\otimes}\mathcal{O}_{d})^{H},\\
& Spec(\mathfrak{Z}(\mathfrak{A}^{d}\underset{\mathcal{O}_{d}^{G}}{\otimes
}\mathcal{O}_{d})) =G/H.
\end{align}
\end{proposition}
\end{enumerate}

On the basis of the above machinery, we can derive the following isomorphisms
and the reciprocity relations of Frobenius type:
\begin{equation}%
\begin{array}
[c]{ccc}%
\mathcal{T}_{\mathfrak{A}}([\rho_{\eta}],\sigma_{\gamma}) & \simeq &
\mathcal{T}_{\mathfrak{A}^{d}}(\rho_{\eta},\tilde{\sigma}_{\gamma})\\
\!\rotatebox{270}{$\simeq$}  &  & \rotatebox{270}{$\simeq$} \\
Rep_{G}([\eta],\gamma) & \simeq & Rep_{H}(\eta,\gamma\upharpoonright_{H})
\end{array}
.
\end{equation}
In use of this, we see that the condition for $\gamma\in\hat{G}$ to belong to
$Spec(\mathfrak{Z}(\pi_{\eta}\upharpoonright_{\mathfrak{A}}(\mathfrak{A}%
)^{\prime\prime}))$ is equivalent to the condition that $\pi_{\gamma}=\pi
_{0}\circ\sigma_{\gamma}$ is a subrepresentation of $\pi\circ\lbrack\rho
_{\eta}]\upharpoonright_{\mathfrak{A}}$, and hence, that the restriction
$\gamma\upharpoonright_{H}$ of $\gamma$ to $H$ contains $\eta\in\hat{H}$:
\begin{equation}
\gamma\in Spec(\mathfrak{Z}(\pi_{\eta}\upharpoonright_{\mathfrak{A}%
}(\mathfrak{A})^{\prime\prime}))\Longleftrightarrow\gamma\in\hat{G}\text{ and
}\exists\eta_{1}\in RepH\text{ s.t. }\gamma\upharpoonright_{H}=\eta\oplus
\eta_{1},
\end{equation}
which is also equivalent to the condition that $\exists v_{1},v_{2}$:
isometries $\in\mathfrak{A}^{d}$ s.t. $\tilde{\sigma}_{\gamma}(B)=v_{1}%
\rho_{\eta}(B)v_{1}^{\ast}+v_{2}\rho_{\eta_{1}}(B)v_{2}^{\ast}$ for
$B\in\mathfrak{A}^{d}$ and $\pi_{0}\circ\sigma_{\gamma}(\mathfrak{A}%
)\subset\pi_{0}\circ\lbrack\rho_{\eta}](\mathfrak{A})$.

\subsection{Interpretation of sector structure: degenerate vacua with order
parameters, Goldstone modes and condensates}

Now the physical meaning of our map, $\Psi:\mathfrak{A}^{d}\ni B\longmapsto
\Psi(B)$, \newline $[\Psi(B)](\dot{g},(\eta,\gamma))=(\omega_{0}\circ
\sigma_{\gamma}\circ m_{G/H}\circ\rho_{\eta}\circ m_{H})(\tau_{\dot{g}}(B))]$,
is clear, and its dual,
\begin{align*}
\Psi^{\ast}  & :M_{1}(H\backslash G)\otimes M_{1}(\underset{\chi\in
\hat{H}}{\amalg}\left\{  \gamma\in\hat{G};\exists\eta_{1}\in RepH\text{ s.t.
}\gamma\upharpoonright_{H}=\chi\oplus\eta_{1}\right\}  )\\
& \rightarrow E_{\mathfrak{A}^{d}},
\end{align*}
gives a \textit{c}$\rightarrow$\textit{q channel}, whose inverse $(\Psi^{\ast
})^{-1}$ exists on the states of $\mathfrak{A}^{d}$ selected by the DHR
criterion, as a \textit{q}$\rightarrow$\textit{c channel} to provide the
physical interpretations of such states in terms of the order parameters in
$\dot{g}\in H\backslash G$, $H$-charge $\eta\in\hat{H}$ and $G$-charge
$\gamma\in\hat{G}$ constrained to $\eta$ by the relation $\gamma
\upharpoonright_{H}\succ\eta$. In view of our starting premise of the
observable algebra $\mathfrak{A}$, however, it looks more natural to take
$A\in\mathfrak{A}$ as the argument of $\Psi$, instead of $B\in\mathfrak{A}%
^{d}$. Then, because of $G$-invariance of $A\in\mathfrak{A}=\mathfrak{F}^{G}
$, $\Psi\upharpoonright_{\mathfrak{A}}$ bocomes independent of $\dot{g}\in
H\backslash G$, $[\Psi(A)](\dot{g},(\eta,\gamma))=(\omega_{0}\circ
\sigma_{\gamma}\circ m_{G/H}\circ\rho_{\eta}\circ m_{H})(\tau_{\dot{g}}%
(A))=(\omega_{0}\circ\sigma_{\gamma}\circ m_{G/H}\circ\rho_{\eta})(A) $,
failing to pick up the information on $H\backslash G$.

To settle this matter, we first consider the physical meaning of the obtained
order parameters:

\begin{itemize}
\item[i)] $H\backslash G$ as \textit{order parameters} to parametrize the
degenerate vacua: in the decomposition of the representation space
$\mathfrak{\hat{H}}$ of $\mathfrak{\hat{F}}$ to pure vacuum representations of
$\mathfrak{F}$ in $\mathfrak{H}$, we get the centre $L^{\infty}(H\backslash
G,d\dot{g})$ with the spectrum $H\backslash G$ which parametrizes the
\textbf{degenerate vacua}\textit{\ }with minimum energy $0$ generated by the
SSB of $G$ up to the unbroken remaining $H$. The physical meaning of this
quantity shows up in such forms as the direction of magnetizations in the
Heisenberg ferromagnets, or, as the \textit{Josephson effect} where the
difference of the phases of Cooper pair condensates between adjacent vacua
accross a junction exhibits such eminent physical effects as the
\textit{Josephson current}\textbf{\ }(see, e.g., \cite{Oji98}).
Mathematically, the Mackey induction from $H$ to $G$ is relevant here. (While
the disjointness $\omega\overset{\shortmid}{\circ}(\omega\circ\tau_{g})$
between any pure vacuum $\omega$ of $\mathfrak{F}$ and $\omega\circ\tau_{g}$
for \textit{any }$g\in G$ s.t. $g\notin H$ might puzzle one about possible
discontinuous behaviours of the order parameter $\dot{g}\in H\backslash G$
under $G$, this is unnecessary worry because of the presence of centre
$C(H\backslash G)$ in $\mathfrak{\hat{F}}$ \textit{in the C*-version} on which
$G$ acts continuously.)

\item[ii)] Internal spectrum $\hat{H}$ of \textit{excited states} on a chosen
vacuum parametrized by a fixed $\dot{g}=Hg\in H\backslash G$: in the
representation space $\mathfrak{H}$ of $\mathfrak{F}$, we see the standard
picture of sectors $(\pi_{\eta},\mathfrak{H}_{\eta})$ with respect to
$\mathfrak{A}^{d}$ parametrized by $\eta\in\hat{H}$, which describes the
internal symmetry aspects of excited states in terms of the unbroken $H$.

\item[iii)] Description of SSB dual to i) in terms of \textit{Goldstone
sectors} within $\mathfrak{A}^{d}$: the sectors w.r.t. $\mathfrak{A}^{d}$ need
still be decomposed into the disjoint representations (=$\mathfrak{A}%
$-sectors) of our genuine observables $\mathfrak{A}$, corresponding to the
``diagonalizations'' of $\mathfrak{Z}(\pi_{\eta}\upharpoonright_{\mathfrak{A}%
}(\mathfrak{A})^{\prime\prime})$ in each $\mathfrak{A}^{d}$-sector with
$\eta\in\hat{H}$. If the group $G$ of spontaneously broken symmetry is
compact, this process is controlled by the \textbf{finite-dimensional
induction/reduction} between $H$ and $G$ on the algebra $\mathfrak{F}$ and by
its \textbf{dual version with respect to local endomorphisms} of
$\mathfrak{A}$ and of $\mathfrak{A}^{d}$. Since the gap between $\mathfrak{A}%
^{d}$ and $\mathfrak{A}$ is essentially due to the Goldstone modes, this may
be interpreted as an abstract version of the Goldstone and/or low-energy
theorems in the sense that they give a \textit{dual} description of the
degenerate vacua in i), in a virtual way \textit{within} a fixed pure vacuum representation.
\end{itemize}

The remarkable parallel and/or reciprocal relations described above and
depicted below, between [$\mathfrak{\hat{F}}$ vs. $\mathfrak{F}$] and
[$\mathfrak{A}$ vs. $\mathfrak{A}^{d}$], or, between [$\mathfrak{F}$ vs.
$\mathfrak{A}$] and [$\mathfrak{\hat{F}}$ vs. $\mathfrak{A}^{d}$],%
\begin{align*}
\text{ \ \ \ \ \ }  & \mathfrak{\hat{F}}\text{ \ \ \ \ \ \ \ \ \ }%
\overset{H\backslash G}{---}\text{\ \ \ \ \ \ \ \ }\mathfrak{F}\\
& \text{ \ }|^{G}\text{ \ \ \ \ \ \ \ \ \ }\diagup_{H}%
\text{\ \ \ \ \ \ \ \ \ \ \ }|^{G}\text{ }\\
& \mathfrak{A}^{d}\text{ \ \ \ \ \ \ \ }-\underset{G/H}{-}%
-\text{\ \ \ \ \ \ \ }\mathfrak{A}\\
& \text{ \ }|_{\text{ }W_{\eta_{\rho}}\simeq H_{\rho}}%
\text{\ \ \ \ \ \ \ \ \ \ \ \ \ \ \ \ \ \ }|_{V_{\gamma_{\sigma}}\simeq
H_{\sigma}}\\
& \text{\ \ }|\text{\ \ \ \ \ \ \ \ \ \ \ \ \ \ \ \ \ \ \ \ \ \ \ \ \ \ \ \ }%
|\\
& \rho_{\eta}(\mathfrak{A}^{d})\overset{\eta\prec\gamma\upharpoonright_{H}%
}{\underset{[\pi_{\eta}](\mathfrak{A})\supset\pi_{\gamma}(\mathfrak{A})}{---}%
}\sigma_{\gamma}(\mathfrak{A}),
\end{align*}
can naturally be understood from the viewpoint of implicit relevance of the
Jones tower structures \cite{Longo, FrKerl}.

On the basis of the above duality between i) and iii), it will be possible to
recover the information on the $H\backslash G$ dependence of states, within a
pure vacuum representation without going over to the other such, through the
closer analysis of the difference term $\eta_{1}\in RepH$ between
$\gamma\upharpoonright_{H}=\eta\oplus\eta_{1}$ and $\eta\in\hat{H}$, which can
be justified by the structure of $G/H$ involved in the proof \cite{TodaMimura}
of finite-dimensional inductions. The verification of this expectation will
fully justify such an heuristic and physical expression that ``the Goldstone
modes search the degenerate vacua in a virtual way''.

While it would be highly interesting to verify it rigorously along the above
iii), we here simply take $\mathfrak{A}^{d}$ as our \textit{extended
observables}. This standpoint is also fully justified by the natural physical
meaning of $\mathfrak{A}^{d}$ as the \textit{maximal local net generated by
the original net }$\mathfrak{A}$, which is just the version, adapted to the
observable net, of the notion of the Borchers classes \cite{Borch60}
consisting of all the \textit{relatively local} fields to absorb the
\textit{arbitrariness in the ``interpolating fields'' }\cite{Nishi}: in spite
of their $G$-non-invariance property, the \textit{Goldstone modes} related to
the homogeneous space $H\backslash G$ are allowed to appear here with the
qualification of such extended observables belonging to $\mathfrak{A}^{d}$ and
they detect the information concerning the position of a pure vacuum
$\dot{g}\in H\backslash G$ among the degenerate vacua, as is exhibited through
the $\dot{g}$-dependence of $\Psi(B)$ for $B\in\mathfrak{A}^{d}$. To be more
precise, one needs to be careful about the distinctions and mutual relations
among the following four levels involving Goldstone modes and order parameters:

\begin{itemize}
\item[i)] degenerate vacua as continuous sectors parametrized by the
\textit{order parameters} $\dot{g}\in H\backslash G$ which is a global notion,

\item[ii)] \textit{Goldstone modes} and \textit{Goldstone sectors} belonging,
respectively, to $\mathfrak{A}^{d}$ and its representation space
$\mathfrak{H}_{\gamma}$, whose massless spectrum is responsible for the
validity of Goldstone theorem,

\item[iii)] \textit{Goldstone multiplet} belonging to $\mathfrak{F}$ and
consisting of Goldstone modes and of condensates responsible for the above i);
this field multiplet transforms under $G$ according to a \textit{linear}
representation, which is nothing but a ``\textit{linear representation of a
homogeneous space}'' according to the definition of \cite{IwaSug}. What is
most confusing is the mutual relation between the Goldstone modes and the
condensates; in the simplest example of SSB (e.g., Heisenberg ferromagnet)
from $G=SO(3)$ to $H=SO(2)$ with $H\backslash G=S^{2}$, a pure vacuum among
degenerate vacua is parametrized and geometrically depicted by a point $p\in
S^{2}$, a condensate by a radius from the centre of the unit ball to $p$, and
the Goldstone modes geometrically expressed by \textit{tangent vectors} at $p$
\textit{tangential to} $S^{2}$ and \textit{orthogonal to the condensate}. The
Goldstone multiplet is an entity in $\mathfrak{F}$ which is behaving as a
three-dimensional covariant vector under $SO(3)$.

\item[iv)] There is a useful physical notion called ``nonlinear realization''
of Goldstone bosons \cite{CallaColeWess}, expressing the above situation in a
nice geometric way and serving as very effective tools in the derivation of
the so-called low energy theorems, such as the soft-pion theorem, to describe
the low energy scattering processes involving Goldstone bosons associated with
SSB. While its functional role is very akin to our Goldstone modes in ii), it
may not be so straightforward to accommodate it literally into the present
context, because of the nonlinear transformation law exhibited in its
transformation property under $G$.
\end{itemize}

As already remarked, we have recourse in the above considerations to the
compactness assumption of the spontaneously broken $G$, which does not hold on
the general ground, as shown in \cite{BDLR92}. While we have benefited from
this assumption in drawing a detailed and concrete picture of the
fiber-structure involved in $Spec(\oplus_{\eta\in\hat{H}}\mathfrak{Z}%
(\pi_{\eta}\upharpoonright_{\mathfrak{A}}(\mathfrak{A})^{\prime\prime}))$, the
essential feature can be expected to survive without this assumption. For
instance, the relations
\begin{equation}
\mathfrak{F}=\mathfrak{A}\rtimes_{\delta}G,\text{ \ \ \ }\mathfrak{A}%
^{d}=\mathfrak{F}^{H}=\mathfrak{A}\rtimes_{\delta}(H\backslash G),
\end{equation}
with a co-action $\delta$ of $G$ on $\mathfrak{A}=\mathfrak{F}^{G}$ are known
to hold in their W*-versions for any locally compact group $G$ \cite{Nak-Take}%
. So the verification of their C*-versions will support the validity of the
essential points of our picture described above that the Goldstone degrees of
freedom related to $H\backslash G$ are contained in $\mathfrak{A}^{d}$ to
describe the SSB-sector structure within the pure vacuum representations,
according to the above iii), etc.

\section{Operational meanings of selection criteria in quantum measurements}

\subsection{Spectral decomposition and probabilistic interpretation}

In view of the importance of the interpretations above, we pick up some
relevant points here from the quantum measurement processes, in regard to the
following basic points:

i) The operator-theoretical notion of spectral decomposition of a self-adjoint
observable $A$ to be measured is equivalent to the algebraic homomorphism
(so-called the map of ``functional calculus''):
\begin{align}
\hat{A} :L^{\infty}(Spec(A))\ni & f \mapsto\ \hat{A}(f)=f(A)\nonumber\\
& :={\int}_{a\in Spec(A)} f(a)\ E_{A}(da)\in\mathfrak{A}^{\prime\prime}\subset
B(\mathfrak{H}),
\end{align}
where $\mathfrak{H}$ is the Hilbert space of the defining representation of
the observable algebra $\mathfrak{A}$ to which our observable $A$ belongs.
Here we omit the symbol for discriminating the original C*-algebra
$\mathfrak{A}$ and its representation in $\mathfrak{H}$, and hence, we will
freely move between C*- and W*-versions without explicit mention. This fits
quite well to the common situations of discussing measurements owing to the
absence of disjoint representations in the purely quantum side $\mathfrak{A}$
with \textit{finite} degrees of freedom (due to Stone-von Neumann theorem). In
such cases, the non-trivial existence of a centre comes only from the
classical system coupled to quantum one (, the former of which need to be
derived from the quantum system with infinite degrees of freedom at the
``ultimate'' levels, though).

ii) To give this homomorphism $\hat{A}$ is (almost) equivalent to giving a
spectral measure $E_{A}$ by
\begin{equation}
E_{A}:\mathcal{B}(Spec(A))\ni\Delta\mapsto\ E_{A}(\Delta):=\hat{A}(\chi
_{\Delta})=\chi_{\Delta}(A)\ \in\ \mathrm{Proj}(\mathfrak{H}),
\end{equation}
on the $\sigma$-algebra $\mathcal{B}(Spec(A))$ on $Spec(A)$ of Borel sets
$\Delta$, identified with the indicator function $\chi_{\Delta}$, taking
values in the set $\mathrm{Proj}(\mathfrak{H})$ of orthogonal projections in
$\mathfrak{H}$. Then the dual map $\hat{A}^{\ast}$ defines a mapping from a
quantum state $\omega$ to a \textit{probability distribution}, $p^{A}%
(\cdot|\ \omega):\mathcal{B}(Spec(A))\ni\ \Delta\ \mapsto\ p^{A}(\Delta
|\omega)=\text{Prob}(A\in\Delta\ |\ \omega):=\omega(E_{A}(\Delta))$, of
measured values in the measurements of $A$ performed in the state $\omega$.
The above reservation ``(almost) equivalent'' is due to the fact that the
reverse direction from a probability distribution to a spectral decomposition
admits a sligthly more general notion, positive-operator valued measure (POM),
which corresponds to a unital completely positive map instead of a
homomorphism and which becomes relevant for treating the set of mutually
non-commutative observables. In any case, the operational meaning of the
mathematical notion of spectral decomposition is exhibited by this
$\hat{A}^{\ast}$ (or, the dual of POM) as a simplest sort of \textit{q}%
$\rightarrow$\textit{c channel} providing the familiar probabilistic interpretation.

iii) To implement physically the spectral decomposition, however, we need some
\textit{physical interaction processes} between the system and the apparatus
through the coupling term of the observable $A\in\mathfrak{A}$ to be measured
and an external field $J$ belonging to the apparatus. While one of the most
polemic issues in the measurement theory is as to how this ``contraction of
wave packets'' is realized consistently with the ``standard'' formulaion of
quantum theory, we here avoid this issue, simply taking such a
``phenomenological'' standpoint that our purpose will be attained if the
composiste system consisting of the object system and the classical system
involving $J$ is effectively (Fourier- or Legendre-) transformed through this
coupled dynamical process into $\mathfrak{A}\otimes C^{\ast}%
\{A\}=:\mathfrak{A}_{A}=C(Spec(A),\mathfrak{A})$, the centre of which is just
the commutative C*-algebra $C^{\ast}\{A\}\simeq C(Spec(A))$ generated by a
self-adjoint operator $A$: $C^{\ast}\{A\}\overset{\iota}{\hookrightarrow}$
$\mathfrak{Z}(\mathfrak{A}_{A})\hookrightarrow\mathfrak{A}_{A}$. So the
\textit{superselection structure} comes in here with sectors parametrized by
the spectrum of the observable $A$ to be measured. (It was the important
contribution of Machida and Namiki \cite{MN} that shed a new light on the
notion of continuous superselection rules, where the focus was, unfortunately,
upon sectors related to \textit{irrelevant unobservable }variables, in sharp
contrast to those discussed here.)

\subsection{Measurement scheme and its realizability}

Then the basic measurement scheme \cite{Ozawa} reduces to the requirement that
all the information on the probability distribution in ii) should be recorded
in and can be read out from this classical part $\{A\}^{\prime\prime
}=L^{\infty}(Spec(A))$ as a mathematical representative of the measuring
apparatus:
\begin{equation}
\omega(E_{A}(\Delta))=p^{A}(\Delta|\omega)=(\omega\otimes\mu_{0})[\hat{\tau
}(\mathbf{1}\otimes\chi_{\Delta})],\label{MeasSchm}%
\end{equation}
where $\mu_{0}$ is some initial state of $\{A\}^{\prime\prime}$ and $\hat
{\tau}\in Aut(\mathfrak{A}_{A})$ describes the effects of dynamics of the
composite system of $\mathfrak{A}$ and $C^{\ast}\{A\}$ (or, more generally, a
dissipative dynamics of a completely positive map also to be allowed).

We are interested here in examining how the problem of a selection criterion
according to our general formulation becomes relevant to the present context.
Applying to any state $\hat{\omega}\in E_{\mathfrak{A}_{A}}$ the uniquely
determined \textit{central decomposition}, we have
\begin{equation}
\hat{\omega}=\int_{Spec(A)}d\mu(a)(\omega_{a}\text{ }\otimes\delta
_{a}),\label{centrl_dec}%
\end{equation}
with some family of states $\{\omega_{a}\}\subset E_{\mathfrak{A}}$ (which are
universally chosen by $\omega_{a}(B):=\langle\psi_{a}\ |\ B\psi_{a}\rangle$
with $A\psi_{a}=a\psi_{a}$ \textit{if} $A$ has only \textit{discrete
spectrum}). What plays important roles here is the instrument $\mathcal{J}%
_{A,\tau}$ \cite{Davies-Lewis} depending on $A\in\mathfrak{A}$ and on the
composiste-system dynamics $\hat{\tau}$ defined by
\begin{align}
\mathcal{I}_{A,\hat{\tau}}:\mathfrak{A}_{A}\ni\hat{B}\longmapsto
\mathcal{I}_{A,\hat{\tau}}(\hat{B})  & :=\int d\mu_{0}(a)(\hat{\tau}(\hat
{B}))(a)\nonumber\\
& =\int d\mu_{0}(a)\delta_{a}(\hat{\tau}(\hat{B}))\in\mathfrak{A,}\\
\mathcal{J}_{A,\hat{\tau}}(\Delta|\omega)(B) & := [\mathcal{I}_{A,\hat{\tau}%
}^{\ast}(\omega)](B\otimes\chi_{\Delta}))=\omega(\mathcal{I}_{A,\hat{\tau}%
}(B\otimes\chi_{\Delta}))\nonumber\\
& =(\omega\otimes\mu)[\hat{\tau}(B\otimes\chi_{\Delta})].
\end{align}
In terms of these notions, Eq. (\ref{MeasSchm}) can be rewritten as
\begin{align}
\hat{A}^{\ast}(\omega)  & =(\mathcal{I}_{A,\hat{\tau}}\circ\iota^{\prime
})^{\ast}(\omega)\nonumber\\
\Longrightarrow\hat{A}^{\ast}  & =\iota^{\prime}{}^{\ast}\circ\mathcal{I}%
_{A,\hat{\tau}}^{\ast},\label{select}%
\end{align}
where $\iota^{\prime\ast}:E_{\mathfrak{A}_{A}^{\prime\prime}}\rightarrow
M_{1}(Spec(A))$ defined by the dual of
\begin{equation}
\iota^{\prime}:\{A\}^{\prime\prime}\ni f\longmapsto\mathbf{1}\otimes
f\in\mathfrak{A}_{A}^{\prime\prime}%
\end{equation}
is the standard (tautological) \textit{q}$\rightarrow$\textit{c channel }to
allow the data read-out from the system-apparatus composite system.
Eq.(\ref{select}) selects out an observable $A$(, or its corresponding
\textit{q}$\rightarrow$\textit{c }channel\textit{\ }$\hat{A}^{\ast}$
describing the probabilistic interpretation of $A$) according to a criterion
as to whether it can be factorized into the standard tautological
\textit{q}$\rightarrow$\textit{c channel }$\iota^{\prime\ast}$ and some
instrument $\mathcal{I}_{A,\hat{\tau}}$. In view of the formal similarity
between the DHR criterion, $\omega=\omega_{0}\circ\rho$, and Eq.(\ref{select}%
), it is interesting to note that what are examined here is \textit{q}%
$\rightarrow$\textit{c }channels, $\hat{A}^{\ast}$ and $\iota^{\prime}{}%
^{\ast}$, the latter of which is a fixed standard one. This criterion is just
for examining whether the measurement of $A$ can actually be materialized by
means of the coupling $\hat{\tau}$ between the system containing $A$ and some
measuring apparatus constituting the composite system $\mathfrak{A}%
_{A}=\mathfrak{A}\otimes\{A\}^{\prime\prime}$. In this sense, the criterion
examines the \textit{realization problem }in the context of control theory
\cite{ArbibMane}, asking whether a suitable choice of an apparatus and a
choice of dynamical coupling can correctly describe the input-output behaviour
of the system.\textit{\ }Once this criterion is valid, its experimental
observation is most conveniently described by the instrument $\mathcal{J}%
_{A,\hat{\tau}}(\Delta|\omega)(B)$ whose interpretation is given \cite{Ozawa} by

1) the probability distribution of the measured value of $A$ in a state
$\omega$ is given by $\mathcal{J}_{A,\hat{\tau}}(\Delta|\omega)(\mathbf{1}%
)=p_{A}(\Delta|\omega),$

2) the final state realized (in the repeatable measurement) after the readout
$a\in\Delta$ is given by $\frac{\mathcal{J}_{A,\hat{\tau}}(da|\omega)}%
{p_{A}(da|\omega)}$,

3) in combination of 1) and 2), the quantity $\mathcal{J}_{A,\hat{\tau}%
}(\Delta|\omega)(B)$ itself can be regarded as the expectation value of
another observable $B\in\mathfrak{A}$ when the initial state $\omega$ goes
into some final state whose $A$-values belong to $\Delta(\subset Spec(A))$.

\subsection{Problem of state preparation as reachability problem}

In the related context, we need to examine the problem of
\textit{reachability} to ask whether there is a controlled way to drive the
(composite) system to any desired state starting from some initial state; this
is nothing but the problem of \textbf{state preparation}, which has not been
seriously discussed, in spite of its vital importance in the physical
interpretation of quantum theory.

For this purpose, we need to define the \textit{c}$\rightarrow$\textit{q
channel} relevant to it. Fixing a family $(\omega_{a})_{a\in Spec(A)}=:\phi$
of states on $\mathfrak{A}$ appearing in the central decomposition
(\ref{centrl_dec}), we can define a \textit{c}$\rightarrow$\textit{q channel}
by
\begin{equation}
C_{A,\phi}:\mathfrak{A}_{A}\ni\hat{B}\longmapsto(Spec(A)\ni a\longmapsto
\omega_{a}(\hat{B}(a)))\in C(Spec(A)),
\end{equation}
and hence, $C_{A,\phi}^{\ast}:M_{1}(Spec(A))\ni\rho\longmapsto C_{A,\phi
}^{\ast}(\rho)\in E_{\mathfrak{A}_{A}}$, where
\begin{align}
C_{A,\phi}^{\ast}(\rho)(\hat{B})  & =\rho(C_{A,\phi}(\hat{B}))=\int
d\rho(a)\omega_{a}(\hat{B}(a))=\int d\rho(a)(\omega_{a}\otimes\delta_{a}%
)(\hat{B}),\nonumber\\
\text{or, }C_{A,\phi}^{\ast}(\rho)  & =\int d\rho(a)(\omega_{a}\otimes
\delta_{a}).
\end{align}
In terms of these, the reachability (or, preparability) criterion can be
formulated as to examine the validity of
\begin{equation}
\omega=\lim_{t\rightarrow\infty}(\iota^{\ast}\circ C_{A,\phi}^{\ast}%
)(\mu_{\hat{\tau}_{t}}),\label{prepare}%
\end{equation}
where $\iota^{\ast}:E_{\mathfrak{A}_{A}}\rightarrow E_{\mathfrak{A}}$ is the
dual of $\iota:\mathfrak{A}\ni B\longmapsto B\otimes\mathbf{1}\in
\mathfrak{A}_{A}$, and the measure $\mu_{\hat{\tau}_{t}}^{\omega}\in
M_{1}(Spec(A))$ is defined through the central decomposition of $(\omega
\otimes\mu_{0})\circ\hat{\tau}_{t}=\int d\mu_{\hat{\tau}_{t}}^{\omega
}(a)\omega_{a}\otimes\delta_{a}$ valid for such an observable $A$ as with
discrete spectrum. If we can find such a suitable coupled dynamics $\hat{\tau
}_{t}$ and an initial and final probability measures $\mu_{0},\mu_{1}\in$
$M_{1}(Spec(A))$ that $\lim_{t\rightarrow\infty}(\omega\otimes\mu_{0}%
)\circ\hat{\tau}_{t}(B\otimes\mathbf{1})=(\omega\otimes\mu_{1})(B\otimes
\mathbf{1})$ for each $B\in\mathfrak{A}$, then a state $\omega$ can actually
be prepared:
\begin{align}
& (\iota^{\ast}\circ C_{A,\phi}^{\ast})(\mu_{\hat{\tau}})(B) =\mu_{\hat{\tau
}_{t}}(C_{A,\phi}(B\otimes\mathbf{1}))=\int d\mu_{\hat{\tau}_{t}}(a)\omega
_{a}(B\otimes\mathbf{1})\nonumber\\
& =(\omega\otimes\mu_{0})\circ\hat{\tau}_{t}(B\otimes\mathbf{1})\underset
{t\rightarrow\infty}{\rightarrow}(\omega\otimes\mu_{1})(B\otimes
\mathbf{1})=\omega(B),
\end{align}
in the sense that there is some operational means specified in terms of
$A\in\mathfrak{A}$, a coupled dynamics $\hat{\tau}_{t}$ and an initial and
final probability measures $\mu_{0},\mu_{1}\in$ $M_{1}(Spec(A))$.

Here, the assumption of discreteness of the spectrum of $A$ is no problem,
since $A$ plays here only a subsidiary role. However, this problem becomes
crucial when we start to examine the \textit{repeatability} of the measurement
of the observable $A$ itself. We compare the above \textit{q}$\rightarrow
$\textit{c channel} $(C_{A,\phi}^{\ast})^{-1}$ with another natural
\textit{q}$\rightarrow$\textit{c channel} $(\iota\circ\hat{A})^{\ast}$, which
can be defined on all the states $\in E_{\mathfrak{A}_{A}}$, independently of
a specific choice of a family $\phi=(\omega_{a})_{a\in Spec(A)}$ of states on
$\mathfrak{A}$, simply as the dual of the composed embedding maps,
$C(Spec(A))\overset{\hat{A}}{\hookrightarrow}\mathfrak{A}\overset{\iota
}{\hookrightarrow}\mathfrak{A}_{A}$. As is seen from the relation,%

\begin{align}
& (\iota\circ\hat{A})^{\ast}(\int d\mu(a)(\omega_{a}\text{ }\otimes\delta
_{a}))(f)\nonumber\\
& =\int d\mu(a)(\omega_{a}\text{ }\otimes\delta_{a}))((\iota\circ
\hat{A})(f))=\int d\mu(a)(\omega_{a}\text{ }\otimes\delta_{a}))(f(A)\otimes
\mathbf{1})\nonumber\\
& =\int d\mu(a)\omega_{a}(f(A))=\int d\mu(a)\int\omega_{a}(dE_{A}(b))f(b),
\end{align}
$(\iota\circ\hat{A})^{\ast}$ is, in general, not equal to $(C_{A,\phi}^{\ast
})^{-1}$, nor has a simple interpretation. If we can choose such a family
$(\omega_{a})_{a\in Spec(A)}$ that $\int f(b)\omega_{a}(dE_{A}(b))=f(a)$ for
$\forall f\in C(Spec(A))$, or equivalently, $\omega_{a}(E_{A}(\Delta
))=\chi_{\Delta}(a)$ for $\forall\Delta$: measurable subset of $Spec(A)$, we
can attain the equality between $(C_{A,\phi}^{\ast})^{-1}$ and $(\iota
\circ\hat{A})^{\ast}$ on the image of $C_{A,\phi}^{\ast}$ in $E_{\mathfrak{A}%
_{A}}$, which can be extended to the whole $E_{\mathfrak{A}_{A}}$ by the use
of the Hahn-Banach extension. As a result, we can attain universally the state
preparations and physical interpretations (in relation to $A$), independently
of a specific choice of the above family $(\omega_{a})_{a\in Spec(A)}$. While
such a choice is always possible for observables $A$ with discrete spectrum,
its impossibility for those $A$ with \textit{continuous spectra} forces us to
consider the \textit{approximate measurement scheme} (see \cite{Ozawa}), which
involves the essential dependence on the choice of the family $(\omega
_{a})_{a\in Spec(A)}$ and the selection of and restriction to preparable and
interpretable states.

In this way, we have seen that this approach provides a simple unified scheme
based upon instruments and channels for discussing various aspects in the
measurement processes without being trapped in the depth of philosophical
issues. So, it will be worthwhile to attempt the possible extension of the
measurement scheme to more general situations involving QFT. It will be also
interesting to examine the problems of state correlations in entanglements, of
state estimation, and so on, in use of the notions of mutual entropy, channel
capacities \cite{Ohya}, Cram\'{e}r-Rao bounds, etc.

Through the above relation with the spectral decomposition of an observable
$A$ and the superselection sectors parametrized by $a\in Spec(A)$, we can
reconfirm the naturalility of our extending the meaning of sectors from their
traditional version of discrete one, to the present version including both: in
SSB, order parameters of continuous family of disjoint states (of
$\mathfrak{A}$) parametrized by $H\backslash G$ and in thermal situations,
(inverse) temperatures $\beta$[=$(\beta^{\mu})$] discriminating pure
thermodynamic phases corresponding also to disjoint KMS states (of
$\mathfrak{A}$), and variety of non-equilibrium local states (\cite{BOR01}).
Our way of unifying these various cases is seen to be quite similar to the
unified treatment of discrete and continuous spectra of self-adjoint operators
in the general theory of spectral decompositions.

\section{Outlook}

We conclude this paper by mentioning some problems under investigation, which
will be reported somewhere.

\begin{itemize}
\item[1.] Treatment of a non-compact group of broken internal symmetry as
remarked in Sec.2.3 and 2.4.

\item[2.] Reformulation of characterization of KMS states: in I;Sec.2 and
I;Sec.3, we have just relied on the known simplicial structure of the set of
all KMS states. To be consistent with the spirit of the present scheme, we
need also to find a version of selection criterion to characterize these KMS
states, whose essence should be found in the \textbf{zeroth law of
thermodynamics }from which the familiar parameter of temperaturs arises (in
combination with the first and second laws in such a form as the passivity
\cite{PuszWoro}). In any case, such a physically interesting problem as
drawing a phase diagram just belongs to the analysis of selection structure in
the present context.

\item[3.] To substantiate the above consideration, it is necessary to develop
a systematic way of treating a chemical potential as one of the order
parameters to be added to temperature. This requires the local and systematic
treatment of conserved currents such as $T_{\mu\nu}$ and $j_{\mu}$, extended
to thermal situations just in a parallel way to the local thermal observables
in \cite{BOR01}.

\item[4.] It would be worthwhile to examine whether the notion of a field
algebra $\mathfrak{F}$ is a simple mathematical device, convenient for making
the interpretation easier from the viewpoint laid out by Klein's Erlangen
programme and no more than that.
\end{itemize}

\end{document}